\documentclass[a4paper, aps, amsmath, 10pt, nofootinbib]{revtex4-2}

\usepackage{tikz}
\usepackage{physics,caption,braket}
\usepackage{xcolor}
\usepackage{widebar,mathpazo}
\usepackage{amsfonts}
\usepackage{hyperref}
\usepackage[normalem]{ulem}

\usetikzlibrary{arrows.meta}

\renewcommand{\bar}{\widebar}

\begin{document}

\title{Scalar field theory under Robin boundary conditions: two-point function and energy-momentum tensor}
\author{David Dudal}
\email{david.dudal@kuleuven.be}
\affiliation{KU Leuven Campus Kortrijk -- Kulak, Department of Physics, Etienne Sabbelaan 53 bus 7657, 8500 Kortrijk, Belgium}

\author{Aaron Gobeyn}
\email{aaron.gobeyn@tu-darmstadt.de}
\affiliation{Technische Universit\"at Darmstadt, Institute for Accelerator Science and Electromagnetic Fields (TEMF), Schlossgartenstr. 8, 64289 Darmstadt,
Germany}

\author{Bruno W.~Mintz}
\email{brunomintz@gmail.com}
\affiliation{UERJ -- Universidade do Estado do Rio de Janeiro,	Instituto de Física -- Departamento de Física Teórica -- Rua São Francisco Xavier 524, 20550-013, Maracanã, Rio de Janeiro, Brazil}

\author{Thomas Oosthuyse}
\email{thomas.oosthuyse@kuleuven.be}
\affiliation{KU Leuven Campus Kortrijk -- Kulak, Department of Physics, Etienne Sabbelaan 53 bus 7657, 8500 Kortrijk, Belgium}

\author{Sebbe Stouten}
\email{sebbe.stouten@kuleuven.be}
\affiliation{KU Leuven Campus Kortrijk -- Kulak, Department of Physics, Etienne Sabbelaan 53 bus 7657, 8500 Kortrijk, Belgium}

\begin{abstract}
    We reconsider four-dimensional scalar field theory in presence of Robin boundary conditions on two parallel plates. These boundary conditions are directly imposed in the path integral definition of the theory via auxiliary fields living on the plates. We discuss how this leads to boundary corrections to the standard energy momentum tensor operator. Via a dimensional reduction to an effective three-dimensional boundary theory, we compute the Casimir energy in terms of the plate separation and the two Robin parameters, as well as the scalar field propagator in the presence of the plates. Coincidentally, the boundary contribution vanishes in the expectation value for the vacuum energy, thereby giving results in full accordance with other energy expressions in the literature for the same setup. We also discuss for which values of the Robin parameters this energy is real-valued.
\end{abstract}
\maketitle
\date{}

\section{Introduction}
    The Casimir effect \cite{Casimir1948influence,Casimir1948TwoPlates}, is renowned as a macroscopically measurable evidence of the rich dynamics of the quantum vacuum, demonstrated both theoretically and experimentally, see e.g. \cite{Bimonte2021measurement} for a
    recent overview and \cite{Plunien:1986ca,Bordag:2001qi,Bordag:2009zz,Milton:2004ya} for more background. Recent interest in the Casimir effect has increased due to advancements in micro-mechanical systems reaching nanometer scales, where the Casimir force becomes significant \cite{Zou2013casimir}. Moreover, lattice simulations of Yang-Mills models suggest an intriguing connection between the Casimir effect and non-perturbative QCD, in particular in relation to a dynamically generated mass scale and boundary dynamics \cite{Chernodub2018casimir,Chernodub:2023dok}, see also \cite{Asorey:2024vaw,Karabali:2018ael}.

    In the QED case, the Casimir effect is usually attributed to virtual particles swarming between uncharged, parallel plates, but it can also be interpreted as a consequence of boundary conditions imposed on quantized electromagnetic field modes. In the realm of functional quantum field theory, this translates into a constraint on the functional integral measure, which can be reformulated directly at the level of the action itself \cite{Bordag:1983zk, Dudal:2020yah} using auxiliary fields living on the boundary surfaces. In the case of gauge theories, this allows one to keep perfect track of issues of gauge invariance, among others, and it allows for full access to the well-developed integration machinery of functional quantum field theory, including the case when interactions are added to the game. It is thus a worthy alternative to the more canonical approach in terms of summing over modes \cite{Bordag:2001qi,Bordag:2009zz}.

    From the action of the theory, one can in principle directly compute the vacuum energy, and thence Casimir energy, via the usual Feynman vacuum bubble diagram expansion. In this paper, we will first follow the procedure outlined in \cite{Dudal:2020yah} and construct the pure boundary effective action for massless real scalar field theory subject to general Robin boundary conditions. Within this effective theory, we find an expression for the scalar field propagator between plates and we also find the perhaps less familiar result that the energy-momentum tensor in the functional formalism also receives boundary corrections.

    The outline of the paper is as follows. In Sec. \ref{sec:functional-integral}, we apply the boundary effective theory method to our problem and calculate both the Casimir energy, as well as the scalar field propagator under Robin (mixed) boundary conditions of this configuration. In Sec. \ref{sec:T00},
    we revisit the calculation of the Casimir energy using the $T^{00}$ component of the energy-momentum tensor, using the formalism of the previous section. That is, we relate the Casimir energy with the auxiliary field that implements the boundary condition. This approach is useful to clarify the role of the energy density on the boundary on the Casimir energy. In the final section, we present our conclusions.


\section{Casimir energy and scalar field propagator between two parallel stationary plates with Robin boundary conditions: path integral method}\label{sec:functional-integral}

	We start from massless Euclidean scalar field theory in $d=4$ spacetime dimensions,
 	\begin{equation}\label{eq:action}
 		S_E = \int \dd^4 x \, \left(\frac{1}{2} \delta_{\mu \nu} \partial_{\mu} \phi \partial_{\nu} \phi \right) =  \int \dd^4 x \, \left\{\frac{1}{2}  \phi (-\partial^2) \phi \right\},
 	\end{equation}
 where $\partial^2$ is the four-dimensional Laplacian operator. Note that we assume that the fields vanish at infinity, $\lim_{|x|\rightarrow\infty}\phi(x)=0$, so that the boundary term in (\ref{eq:action}) vanishes. In what follows, we will write the position vector as $x=(x_0,x_1,x_2,x_3)$.

Now, to set up the Casimir problem, we consider two parallel ``plates'', one located at $x_3 = z = L/2$ and another one located at $x_3 = z = -L/2$. The boundary condition we impose on the real scalar field $\phi(x)$ at these plates will be of the Robin type, that is, an interpolation between Dirichlet and Neumann conditions parameterized by two constants $g$ and $\bar{g}$ (one for each plate) with dimensions of inverse length, so that
        \begin{equation}\label{robin}
 		\begin{cases}
 			&\frac{\partial}{\partial n} \phi - g \phi = 0 \quad \text{at $z = L/2$ plate}, \\
 			&\frac{\partial}{\partial n} \phi - \bar{g} \phi = 0 \quad \text{at $z = -L/2$ plate.}
  		\end{cases}
        \end{equation}
Here, $\partial/\partial n$ denotes the normal derivative. For the $z=L/2$ plate, $n_{\mu} = (0,0,0,-1)$, so that $\partial/\partial n = -\partial_z$. For the $z=-L/2$ plate, $n_{\mu} = (0,0,0,1)$, so that $\partial/\partial n=\partial_z$.

We can incorporate the boundary condition \eqref{robin} into the functional integral via appropriate $\delta$-functions \cite{Bordag:1983zk}. The generating functional is given by
\begin{eqnarray}
 Z[J]=\int[D\phi]\delta\left(\frac{\partial\phi}{\partial z}+g\phi\right)_{z=L/2}   \delta\left(\frac{\partial\phi}{\partial z}-\bar{g}\phi\right)_{z=-L/2} \exp\left(-S_E+\int d^4x J\phi\right).
\end{eqnarray}

Note that the arguments of the functional deltas are functions of three coordinates only: $t=x_0$, $x=x_1$ and $y=x_2$. Let us merge those in the three-component Euclidean vector $\vec x=(x_0,x_1,x_2)$. Since the constraints due to the boundary conditions consist of a function of the vector $\vec x$, each of the functional deltas can be exponentiated by using the identity
\begin{eqnarray}
    \delta[{\cal F}] = \int[Db]\exp\left[
    -i\int\,d^3x\,b(\vec x){\cal F}(\vec x)
    \right],
\end{eqnarray}
where $d^3x=dx_0\,dx_1\,dx_2$ and $\cal F$ is the boundary condition (\ref{robin}). This leads to
\begin{eqnarray}
    Z[J]=\int[D\phi][D\bar b][Db]\exp\left[-S_E - S_{bc}+\int d^4x J\phi\right],
\end{eqnarray}
where
\begin{eqnarray}
    S_{bc} = \int d^3x\,dz\left[
    i b(\vec x)\delta\left(z-\frac L2\right)\left(\frac{\partial\phi}{\partial z}(\vec x,z) + g\phi(\vec x,z)\right)
    + i \bar b(\vec x)\delta\left(z+\frac L2\right)
    \left(\frac{\partial\phi}{\partial z}(\vec x,z) - \bar g\phi(\vec x,z)\right)
    \right]
\end{eqnarray}
is the contribution of the boundary conditions to the classical action. Note that one can interpret the $b$ and $\bar b$ fields  as Lagrange multipliers that enforce the boundary conditions on the plates.

 	Next, following \cite{Dudal:2020yah} (see also \cite{Canfora:2022xcx,Oosthuyse:2023mbs} and similar ideas in \cite{Golestanian:1998bx}), we want to integrate out the $\phi$-fields, leaving us with an effective (lower-dimensional) $b$-field action. The easiest way to accomplish this is to work in Fourier (momentum) space. We follow the conventions of \cite{Dudal:2020yah}. The resulting action (with sources) in Fourier space is then given by
  \begin{eqnarray}\label{actievol}
 S_E + S_{bc} - \int J\phi&=& \int \frac{d^3k\,dk_z}{(2 \pi)^4} \left\{
     \frac{1}{2} \phi(-k) k^2 \phi(k)
   + (k_z +i g) e^{-i k_z L/2} b(- \vec{k}) \phi(k)
   + (k_z-i \bar{g}) e^{i k_z L/2} \bar{b}(- \vec{k})\phi(k)
   -J(-k)\phi(k)
   \right\}\nonumber\\
   &=& \int \frac{d^3k\,dk_z}{(2 \pi)^4} \left\{
     \frac{1}{2} \phi(-k) k^2 \phi(k)
   - A_{J\bar bb}(-k)\phi(k)
   \right\}
 \end{eqnarray}
where $k=(\vec k,k_z)$ and we defined
\begin{eqnarray}\label{eq:lambdaJbb}
    A_{J\bar bb}(k):=J(k) + (k_z-ig)e^{ik_zL/2}b(\vec k)
    + (k_z+i\bar g)e^{-ik_zL/2}\bar b(\vec k).
\end{eqnarray}

With the action in this form, the functional integration of the $\phi$ field is straightforward, leading to
\begin{eqnarray}\label{eq:zJ_integration_bbar}
    Z[J] = \left[\det(-\partial^2)\right]^{-1/2}\int[D\bar b][Db]\exp\left[
      \frac12\int\frac{d^3k\,dk_z}{(2\pi)^4}
      A_{J\bar bb}(-k)\frac{1}{\vec k^2+k_z^2}A_{J\bar bb}(k)
    \right]
\end{eqnarray}
In order to calculate the functional integral on the auxiliary fields $b$ and $\bar b$, we note that
\begin{eqnarray}\label{eq:lambda-k2-lambda}
    \frac12\int\frac{d^3k\,dk_z}{(2\pi)^4}
      A_{J\bar bb}(-k)\frac{1}{k^2}A_{J\bar bb}(k) &=&
      \frac12\int \frac{d^4k}{(2\pi)^4} J(-k)\frac{1}{k^2}J(k) - \frac12
      \int \frac{d^3k}{(2\pi)^3} V_i(-\vec k) \mathbb{D}_{ij}(\vec k) V_j(\vec k) + \int \frac{d^3k}{(2\pi)^3} L_i(-\vec k)\,V_i(\vec k)\nonumber\\
\end{eqnarray}
where
\begin{eqnarray}
    V_1(\vec k) &=& ib(\vec k),\nonumber\\
    V_2(\vec k) &=& i\bar b(\vec k),\nonumber\\
\end{eqnarray}
and the matrix that has $\mathbb{D}_{ij}$ as elements is
\begin{eqnarray}\label{eq:Dmatrix}
    \mathbb{D}(\vec k) &=& -\int\frac{dk_z}{2\pi}\frac{1}{\vec k^2+k_z^2}
    		\begin{pmatrix}
			k_z^2 + g^2 & e^{-ik_zL} (ig+k_z)(i\bar g +k_z) \\
			e^{ik_zL} (ig-k_z)(i\bar g -k_z) & k_z^2 + \bar{g}^2
		\end{pmatrix}
              \nonumber\\
              &=&     		
        \frac{1}{2|\vec k|}\begin{pmatrix}
		 |\vec k|^2 - g^2 & (|\vec k|-g)(|\vec k|-\bar{g})e^{-|\vec k|L} \\
		(|\vec k|-g)(|\vec k|-\bar{g})e^{-|\vec k|L}	 &
        |\vec k|^2 - \bar{g}^2
		\end{pmatrix},
\end{eqnarray}

and finally
\begin{eqnarray}\label{eq:sources_L}
    L_1(-\vec k) &=& -i\int\frac{dk_z}{2\pi}J(-\vec k,-k_z)\left(\frac{k_z-ig}{\vec k^2+k_z^2}\right)e^{ik_zL/2}, \nonumber\\
    L_2(-\vec k) &=& -i\int\frac{dk_z}{2\pi}J(-\vec k,-k_z)\left(\frac{k_z+i\bar{g}}{\vec k^2+k_z^2}\right)e^{-ik_zL/2}.
\end{eqnarray}

Note that the integrations that lead to the elements of $\mathbb{D}(\vec k)$ are regularized and listed in the Appendix.
After integrating the auxiliary fields $b$ and $\bar b$, the generating functional then becomes
\begin{eqnarray}\label{eq:bareZ_J}
       Z[J] = \left[\det(-\partial^2)\right]^{-1/2}
       \left[\det(\mathbb{D})\right]^{-1/2}
       \exp\left( \frac12\int \frac{d^4k}{(2\pi)^4} J(-k)\frac{1}{k^2}J(k) +
      \frac12\int\frac{d^3k}{(2\pi)^3}
      L_i(-\vec k)\left[
      \mathbb{D}^{-1}(\vec k)\right]_{ij}L_j(\vec k)
    \right),
\end{eqnarray}
where
\begin{eqnarray}
    \mathbb{D}^{-1}(\vec k) =
    \frac{1}{\det_2\mathbb{D}(\vec k)}
     \begin{pmatrix}
			|\vec k|^2 - \bar{g}^2 & -(|\vec k|-g)(|\vec k|-\bar{g})e^{-|\vec k|L} \\
			-(|\vec k|-g)(|\vec k|-\bar{g})e^{-|\vec k|L} & |\vec k|^2 - g^2
	\end{pmatrix},
\end{eqnarray}
and
\begin{eqnarray}
    {\rm{det}}_2 \mathbb{D}(\vec k) = \frac{(|\vec k|^2-g^2)(|\vec k|^2-\bar{g}^2)-(|\vec k|-g)^2(|\vec k|-\bar{g})^2e^{-2|\vec k|L}}{2|\vec k|}
\end{eqnarray}
is the $2\times2$ determinant of the matrix $\mathbb{D}$, so that the full determinant is given by
\begin{eqnarray}\label{eq:bare_detD}
    \det\mathbb{D}=\det\,\!_k\,\left[\det\,\!_2\mathbb{D}(\vec k)\right] &=& \exp\Tr\,\!_k\ln\left[{\rm{det}}_2 \mathbb{D}(\vec k)\right]
    =\exp\left[L_0L_1L_2\int\frac{d^3k}{(2\pi)^3}
    \ln {\rm{det}}_2 \mathbb{D}(\vec k)   \right]\nonumber\\
    &=& \exp\left[L_0L_1L_2\int\frac{d^3k}{(2\pi)^3}
    \log\frac{(|\vec k|^2-g^2)(|\vec k|^2-\bar{g}^2)-(|\vec k|-g)^2(|\vec k|-\bar{g})^2e^{-2|\vec k|L}}{2|\vec k|}   \right],
\end{eqnarray}
where $L_0$ is the extent of the (Euclidean) time dimension, whereas $L_1$ and $L_2$ correspond to the sides of the plate (which are parallel to the $z=0$ plane). Note that  the plates are taken to be both static and with infinite sides\footnote{Actually, it would be more precise to state that the plates are so close that $L/L_1\ll 1$ and $L/L_2\ll 1$.}, so that $L_i\rightarrow\infty$ $(i=0,1,2)$.

Given expression (\ref{eq:bareZ_J}) for the generating functional, let us now explicitly calculate the Casimir energy and the propagator of the theory under the influence of Robin boundary conditions.

\subsection{Casimir energy}

The integral (\ref{eq:bare_detD}) is not convergent and therefore, some
renormalization prescription is required. We choose, as usual,
to set the vacuum energy to zero when the plates are far apart, i.e.,
\begin{eqnarray}
    \lim_{L\rightarrow\infty} E_{vac}(L) = 0.
\end{eqnarray}

Such a renormalization condition can be implemented
straightforwardly by adding a $L-$independent (and infinite)
counterterm $S_{vac}$ to the action. With such a term,
\begin{eqnarray}
    \det(-\partial^2)\det\mathbb{D}(\vec k) \;\;\longrightarrow\;\;
    \det\mathbb{D}_{ren}(\vec k) &=& \exp\left[L_0L_1L_2
    \int\frac{d^3k}{(2\pi)^3}    \ln\left(
    1-\frac{(|\vec k|-g)^2(|\vec k|-\bar{g})^2e^{-2|\vec k|L}}{(|\vec k|^2-g^2)(|\vec k|^2-\bar{g}^2)} \right)  \right]
\end{eqnarray}
and the renormalized vacuum energy (i.e., the Casimir energy) is finally  given by
\begin{equation}
    E_{Cas}(g,\bar g, L)=-\frac{1}{L_0}\ln Z[J=0] = \frac{L_1L_2}{2}
    \int\frac{d^3k}{(2\pi)^3}    \ln\left(
    1-\frac{(|\vec k|-g)(|\vec k|-\bar{g})e^{-2|\vec k|L}}{(|\vec k|+g)(|\vec k|+\bar{g})} \right)
\end{equation}
where $L_1L_2$ is the area of the plates.

	In order to put the integral in a simpler form, we now go to spherical coordinates. Notice that the angular integrals can immediately be computed which simply gives a factor of $4 \pi$, hence, the Casimir energy per unit area reads
	\begin{equation}
		\frac{E_{Cas}(g,\bar g, L)}{L_1L_2} = \frac{1}{4 \pi^2} \int_0^{\infty} \dd r \, r^2 \left\{ -2Lr + \ln(e^{2Lr} - \frac{(r-g)(r - \bar{g})}{(r+g)(\bar{g} + r)}) \right\}.
 	\end{equation}
 	We can make the $L$-dependence more explicit by considering a substitution $x = rL$, we then find
 	 	\begin{equation}
 		\frac{E_{Cas}(g,\bar g, L)}{L_1L_2} = \frac{1}{4 \pi^2 L^3} \int_0^{\infty} \dd x \, x^2 \left\{ -2x + \ln(e^{2x} - \frac{(x-gL)(x - \bar{g} L)}{(x + gL)(\bar{g}L + x)}) \right\}.
   	\end{equation}
   	 It is useful to define new dimensionless parameters $q = gL$ and $\bar{q} = \bar{g}L$, which gives us the Casimir energy under the form
   \begin{equation}\label{res1}
   		\frac{E_{Cas}(q,\bar q)}{L_1L_2} = \frac{1}{4 \pi^2 L^3} \int_0^{\infty} \dd x \, x^2 \left\{ -2x + \ln(e^{2x} - \frac{(x-q)(x - \bar{q})}{(x + q)(x + \bar{q})}) \right\}
   	\end{equation}
	shown in Figure 1.

\subsubsection{Stability constraint}
The setup considered here was also treated in \cite{deAlbuquerque:2003qbk} using a slightly different computational scheme. Indeed, adapting to our notational conventions, in their path integral the underlying base space of the fields was restricted to $z\in[-L/2,L/2]$, while we consider fields defined over full space $\mathbb{R}^4$. They found
  	\begin{equation}
  		\frac{E_{Cas}^{\text{Alb}}(q,\bar q)}{L_1L_2} = - \frac{\pi^2}{1440 L^3} + \frac{1}{4 \pi^2L^3} \int_0^{\infty} \dd x \, x^2 \ln(1 + \frac{2 (\bar{q}+q) x}{(x+\bar q)(x + q)} \frac{1}{e^{2x} - 1})
  	\end{equation}
 which does equal our result \eqref{res1} upon recombination of the two terms and usage of the standard integral

 \begin{equation}\label{standard}
   \frac{1}{4\pi^2}\int_0^\infty dx  x^2\ln(1-e^{-2x})=-\frac{\pi^2}{1440}
 \end{equation}
 as it appears for example in the pure Dirichlet boundary condition situation, see e.g.~\cite{Ambjorn:1981xw,Plunien:1986ca}. It is interesting to note that also in this case, similar to the QED case \cite{Dudal:2020yah}, the ``outer region'' of the plates is of no consequence whatsoever for the Casimir energy itself.

\begin{figure}[t] 
\centering
\includegraphics[scale=1]{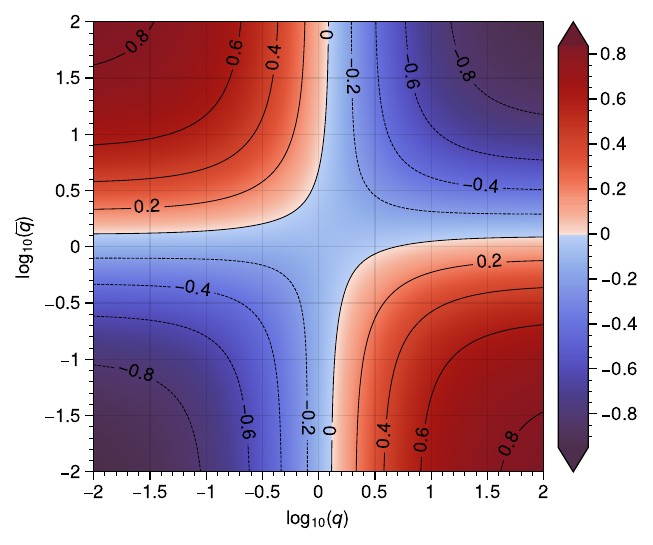}
\caption{The dimensionless Casimir energy \(\frac{1440}{\pi^2 L^3} V_L (q, \bar q)\), expressed in $q = gL$ and $\bar{q} = \bar{g}L$.}
\label{fig}
\end{figure}

To ensure a real-valued energy, the presence of the logarithm in \eqref{res1} dictates that $q\geq 0$ and $\bar q\geq0$ as otherwise there would be at least one single pole in the argument of the logarithm, at $x=-q>0$ and/or $x=-\bar q>0$, where it switches sign. These restrictions $q\geq 0$ and $\bar q\geq0$ were also briefly mentioned in \cite{deAlbuquerque:2003qbk} as to avoid the possible appearance of tachyons.  These constraints can however also be understood from a classical viewpoint. Indeed, the total energy is given by
\begin{equation}\label{energy1}
  E=E_{\tiny bulk}+E_{\tiny boundary},\qquad \frac{dE}{dt}=0,
\end{equation}
with
\begin{equation}\label{energy2}
  E_{\tiny bulk}(t)=\frac{1}{2}\int_{-\infty}^{\infty}dx \int_{-\infty}^{\infty}dy \int_{-L/2}^{L/2}dz \left((\partial_t\phi)^2+(\partial_i\phi)^2\right)
\end{equation}
(where the index $i$ exceptionally runs over $\{x,y,z\}$), so that, upon using the bulk equations of motion and partial integration on the first term,
\begin{eqnarray}\label{energy3}
  \frac{dE_{\tiny bulk}}{dt}&=&\int_{-\infty}^{\infty}dx \int_{-\infty}^{\infty}dy \int_{-L/2}^{L/2}dz \left(\partial_t\phi\partial_i^2\phi+\partial_i\phi \partial_t\partial_i\phi\right)\nonumber\\
  &=& \left[\partial_t \phi\partial_z \phi\right]_{z=-L/2}^{z=L/2}-\int_{-\infty}^{\infty}dx \int_{-\infty}^{\infty}dy \int_{-L/2}^{L/2}dz \left( \partial_i \partial_t\phi \partial_i\phi\right)+\int_{-\infty}^{\infty}dx \int_{-\infty}^{\infty}dy \int_{-L/2}^{L/2}dz \left( \partial_i\phi \partial_t\partial_i\phi\right)\nonumber\\
  &=& \left.\partial_t \phi\partial_z \phi\right|_{z=L/2}-\left.\partial_t \phi\partial_z \phi\right|_{z=-L/2}
\end{eqnarray}
Now, using the Robin conditions \eqref{robin}, this yields
\begin{eqnarray}\label{energy4}
  \frac{dE_{\tiny bulk}}{dt}&=& -\left. g\phi\partial_t \phi\right|_{z=L/2}-\left.\bar g\phi\partial_t \phi \right|_{z=-L/2}\nonumber\\
  &=& -\frac{1}{2}\frac{d}{dt}\left( g\left.\phi^2\right|_{z=L/2}+\bar g \left.\phi^2\right|_{z=-L/2}\right)
\end{eqnarray}
from which we can read off
\begin{eqnarray}\label{energy5}
  E_{boundary}&=& \frac{1}{2}\left( g\left.\phi^2\right|_{z=L/2}+\bar g \left.\phi^2\right|_{z=-L/2}\right)
\end{eqnarray}
As such, to ensure that no infinite amount of energy can flow into the bulk from the boundary, stability requires $E_{boundary}\geq 0$ and thus $g\geq 0$, $\bar g\geq 0$. Note that, in a one-dimensional setting (where the classical field is akin to an ideal string), the Robin boundary condition can be interpreted as the effect of a massless ring attached to the string, and an ideal spring (of elastic constant $g$ or $\bar g$). Thus, the conditions $g\geq 0$ and $\bar g\geq 0$ simply reflect the fact that the boundary exerts a sort of restoring force one the field, temporarily absorbing some energy and returning it completely to the field afterwards. It is interesting to note that a violation of such a stability condition would lead to an imaginary part of the Casimir energy. In a quantized theory this possibility would lead to tachyons, as commented in \cite{deAlbuquerque:2003qbk}.

\subsection{Two-point correlation function}

Let us now focus on the computation of the two-point correlation function for the field $\phi$, under the influence of Robin boundary conditions (\ref{robin}). The generating functional (\ref{eq:bareZ_J}) can be written as
\begin{eqnarray}\label{eq:renormZ_J}
       Z[J] = \exp\left[-L_0\,E_{Cas}(g,\bar{g},L)\right]
       \exp\left( \frac12\int \frac{d^4k}{(2\pi)^4} J(-k)\frac{1}{k^2}J(k) +
      \frac12\int\frac{d^3k}{(2\pi)^3}
      L_i(-\vec k)\left[
      \mathbb{D}^{-1}(\vec k)\right]_{ij}L_j(\vec k)
    \right),
\end{eqnarray}
so that the momentum-space connected two-point function is given by
\begin{eqnarray}
    \vev{\phi(p)\phi(q)} &=& \left.\frac{\delta^2}{\delta J(q)\delta J(p)}\ln Z[J]\right|_{J=0}\nonumber\\
    &=& \frac12\frac{\delta^2}{\delta J(q)\delta J(p)}\left\{\int \frac{d^4k}{(2\pi)^4} J(-k)\frac{1}{k^2}J(k) +
      \int\frac{d^3k}{(2\pi)^3}
      L_i(-\vec k)\left[
      \mathbb{D}^{-1}(\vec k)\right]_{ij}L_j(\vec k)\right\}_{J=0}\nonumber\\
      &=& \frac{\delta(p+q)}{p^2} + D_{Robin}(p,q),
\end{eqnarray}
where
\begin{eqnarray}
    D_{Robin}(p,q)&=&\int\frac{d^3k}{(2\pi)^3}
    \left\{
      \frac{\delta L_i(-\vec k)}{\delta J(p)}
      \left[\mathbb{D}^{-1}(\vec k)\right]_{ij}
      \frac{\delta L_j(\vec k)}{\delta J(q)}
    \right\}_{J=0}.
\end{eqnarray}

Note that we used the fact that the matrix $\mathbb{D}^{-1}(\vec k)$ is both symmetric and an even function of $\vec k$. From the definitions (\ref{eq:sources_L}) for the ``sources'' $L_i$, we find
\begin{eqnarray}
    \frac{\delta L_1(-\vec k)}{\delta J(p)} &=& i\frac{\delta^{(3)}(\vec k+\vec p)}{2\pi}\left(\frac{p_z+ig}{\vec p^2+p_z^2}\right)e^{-ip_zL/2}\nonumber\\
    \frac{\delta L_1(\vec k)}{\delta J(q)} &=& i\frac{\delta^{(3)}(\vec k-\vec q)}{2\pi}\left(\frac{q_z+ig}{\vec q^2+q_z^2}\right)e^{-iq_zL/2}\nonumber\\
    \frac{\delta L_2(-\vec k)}{\delta J(p)} &=& i\frac{\delta^{(3)}(\vec k+\vec p)}{2\pi}\left(\frac{p_z-i\bar{g}}{\vec p^2+p_z^2}\right)e^{ip_zL/2}\nonumber\\
    \frac{\delta L_2(\vec k)}{\delta J(q)} &=& i\frac{\delta^{(3)}(\vec k-\vec q)}{2\pi}\left(\frac{q_z-i\bar{g}}{\vec q^2+q_z^2}\right)e^{iq_zL/2}
\end{eqnarray}

so that
\begin{eqnarray}
    D_{Robin}(p,q)&=&-\frac{1}{(2\pi)^5}\frac{\delta^{(3)}(\vec q+\vec p)}{(|\vec p|^2+p_z^2)(|\vec q|^2+q_z^2)\det_2\mathbb{D}(\vec p)}\left\{
     (|\vec p|^2-\bar{g}^2)(p_z+ig)(q_z+ig)e^{-i(p_z+q_z)L/2}  \right.\nonumber\\
    &&\;\;-(|\vec p|-g)(|\vec p|-\bar{g})e^{-|\vec p|L}\left[(p_z+ig)(q_z-i\bar{g})e^{-i(p_z-q_z)L/2}+(p_z-i\bar{g})(q_z+ig)e^{i(p_z-q_z)L/2}
    \right]+\nonumber\\
     && \left.\;\;\;\;+ (|\vec p|^2-g^2)(p_z-i\bar{g})(q_z-i\bar{g})e^{i(p_z+q_z)L/2}
    \right\}.
\end{eqnarray}

Note that, since the boundary conditions break translational invariance, the momentum-space propagator is not a function of a single momentum variable. As a consequence, the position space propagator
\begin{eqnarray}
    \vev{\phi(x)\phi(y)} = \int\frac{d^4p}{(2\pi)^4}\frac{d^4q}{(2\pi)^4}e^{ip\cdot x+iq\cdot y}\vev{\phi(p)\phi(q)}
\end{eqnarray}
will {\it not} be a function of $x-y$, as in the absence of the plates. However, there is still a residual translational symmetry in the directions orthogonal to the plates, as implied by the presence of $\delta^{(3)}(\vec p+\vec q)$. This is the most general propagator one can find, for different Robin boundary conditions per plate, and as such are a generalization of those reported earlier in literature, \cite{BrownMaclay1969,Barone:2003crf,Albuquerque:1997BJP,Barone:2016rnf}.


\section{Casimir energy via the energy-momentum tensor}\label{sec:T00}

A {\it conditio sine qua non} for any method rooted in the energy-momentum tensor is evidently the usage of the correct tensor. For the current purposes, we only need to concern ourselves with the (integrated) $00$-component of the energy-momentum tensor. Due to the integration, we will not bother to optimize the tensor, as this typically happens via adding appropriate total derivatives to the classical (Noether derived) tensor which are thence of no consequence for our purposes \cite{Belinfante1940OnTC,Baker:2021hly,Freese_2022}. The Noether procedure for translations in the (Euclidean) time direction, applied to the action \eqref{actievol} (but with $J=0$), renders
\begin{equation}\label{tensor1}
  T_{00}=\underbrace{- \partial_0 \phi \partial_0 \phi + \frac{1}{2} \partial_{\mu} \phi \partial_{\mu} \phi}_{T_{00}^{bulk}} + \underbrace{B(\partial_z+g)\phi + \bar B(\partial_z-\bar g)\phi}_{T_{00}^{boundary}},
\end{equation}
where
\begin{eqnarray}
    B(\vec x,z) &=& ib(\vec x)\delta\left(z-\frac L2\right),\nonumber\\
    \bar B(\vec x,z) &=& i\bar{b}(\vec x)\delta\left(z+\frac L2\right).
\end{eqnarray}

We wish to draw particular attention to the last two terms, which are ``pure boundary''. To the best of our knowledge, this kind of contribution is not often discussed when considering computations of the vacuum energy in presence of Casimir boundaries \cite{Plunien:1986ca,Bordag:2009zz}. (An example of analysis of the surface energy can, however, be found in \cite{Romeo:2000wt}, but here we shall take a different approach to this problem.) Perhaps this is no surprise since, as we shall see shortly, the surface terms of $T_{00}$ give a net zero contribution upon taking the integrated expectation value.

Given that the action (\ref{actievol}) is quadratic in the fields, the classical equations of motion are formally equivalent to the Heisenberg equations for the field operators. This implies that the field operators $\phi$, $b$, and $\bar{b}$ are related to each other just like their classical counterparts, i.e.,
\begin{eqnarray}\label{eom1}
    0 &=& \left.\frac{\delta (S_E+S_{bc})}{\delta\phi(k)}\right|_{\phi_{cl}} \nonumber\\
    \Rightarrow \phi_{cl}(k) &=& \frac{1}{k^2}\left[(k_z-ig)e^{ik_zL/2}b(\vec k) + (k_z+i\bar{g})e^{-ik_zL/2}\bar{b}(\vec k)\right]\nonumber\\
    &=& \frac{1}{k^2}\left.A_{J\bar{b}b}(k)\right|_{J=0},
\end{eqnarray}
where $A_{J\bar{b}b}$ was defined in (\ref{eq:lambdaJbb}) and $k^2=\vec k^2+k_z^2$.
This expression can be directly used in $T_{00}^{bulk}$ (now regarding the fields as operators). As a result, the vacuum energy density per unit area for the bulk part reads
\begin{eqnarray}
		\frac{E_{Cas}^{bulk}}{L_1L_2} &=& \int \dd z \, \left\langle T_{00}^{bulk} \right\rangle \nonumber\\
	&=& \int \frac{\dd^4 k}{(2 \pi)^4} \,  \left( \frac{1}{2} k^2 - k_0^2 \right) \left\langle \phi(-k)\phi(k) \right\rangle \nonumber\\
    &=& \int \frac{\dd^4 k}{(2 \pi)^4}
    \frac{\frac12k^2-k_0^2}{k^4}
    \left.\left\langle A_{J\bar{b}b}(-k)A_{J\bar{b}b}(k)\right\rangle\right|_{J=0},
    \end{eqnarray}
 where in the second equality we have used that $\langle \phi(p)\phi(q)\rangle$ is proportional to $\delta^{(3)}(\vec p+\vec q)$. Using (\ref{eq:lambda-k2-lambda}) with $J=0$, we now find
 \begin{eqnarray}\label{eq:E0-two-terms}
     \frac{E_{Cas}^{bulk}}{L_1L_2} &=& - \frac12
      \int \frac{d^3k}{(2\pi)^3} \left\langle V_i(-\vec k) \mathbb{D}_{ij}(\vec k) V_j(\vec k) \right\rangle - \int\frac{d^4k}{(2\pi)^4} \left\langle V_i(-\vec k)\frac{k_0^2}{k^4}\Delta_{ij}(k)V_j(\vec k) \right\rangle,
 \end{eqnarray}
where
\begin{eqnarray}
    \Delta_{ij}(k) = \begin{pmatrix}
				k_z^2 + g^2 & e^{-i k_z L} (k_z+ig)(k_z+i\bar{g}) \\
				e^{i k_z L} (k_z-ig)(k_z-i\bar{g}) & k_z^2 + \bar{g}^2
 			\end{pmatrix}.
\end{eqnarray}

	We have already dealt with the first term of (\ref{eq:E0-two-terms}) in the Sec. \ref{sec:functional-integral}, so only the second one needs some further analysis. Again, we can first integrate out the $k_z$-dependence. We essentially have two integrals to compute	i.e.,
	\begin{equation}
		\begin{split}
			\int \frac{\dd k_z}{2 \pi} \frac{1}{k^4} \Delta_{11}(k)= \frac{1}{4 \abs{\vec{k}}} \left(1 + \frac{g^2}{\abs{\vec{k}}^2} \right),
		\end{split}
	\end{equation}
	and
 \begin{equation}
		\begin{split}
			\int \frac{\dd k_z}{2 \pi} \frac{1}{k^4} \Delta_{12}(k)
			&= \frac{e^{- \abs{\vec{k}} L}}{4 \abs{\vec{k}}} \left\{1 - \abs{\vec{k}} L - \frac{g \bar{g}}{\abs{\vec{k}}^2} (1 + \abs{\vec{k}} L) + (g + \bar{g}) L \right\}.
		\end{split}
	\end{equation}
 We thus find that
		\begin{equation}
		\frac{E_{Cas}^{bulk}}{L_1L_2} = -\int \frac{\dd^3 k}{(2 \pi)^3} \left\langle V_i^T(\vec{k}) \left(\mathbb{H}_{ij}(\vec{k}) + \frac{1}{2}  \mathbb{D}_{ij}(\vec{k}) \right) V_j(- \vec{k}) \right\rangle,
	\end{equation}
where
\begin{equation}
		\mathbb{H}_{ij}(\vec{k}) = \frac{k_0^2}{4 \abs{\vec{k}}}
		\begin{pmatrix}
			1 + \frac{g^2}{\abs{\vec{k}}^2} & e^{- \abs{\vec{k}} L} \left\{1 - \abs{\vec{k}} L - \frac{g \bar{g}}{\abs{\vec{k}}^2} (1 + \abs{\vec{k}} L) + (g + \bar{g}) L \right\} \\
			e^{- \abs{\vec{k}} L} \left\{1 - \abs{\vec{k}} L - \frac{g \bar{g}}{\abs{\vec{k}}^2} (1 + \abs{\vec{k}} L) + (g + \bar{g}) L \right\} & 1 + \frac{\bar{g}^2}{\abs{\vec{k}}^2}
		\end{pmatrix}
	\end{equation}

Using the fact that the two-point correlation function
\begin{eqnarray}
    \left\langle V_i(\vec{k}) V_j(- \vec{k}) \right\rangle = -(\mathbb{D}^{-1})_{ji}(\vec{k}),
\end{eqnarray}
as can be seen by inspecting Eqns.
 (\ref{eq:zJ_integration_bbar}) and (\ref{eq:lambda-k2-lambda}) after taking $J=0$,
we find the following expression for the bulk Casimir energy
 	\begin{equation}\label{bulk1}
 		\frac{E_{Cas}^{bulk}}{L_1L_2} = -\int \frac{\dd^3 k}{(2 \pi)^3} \mathbb{T}_{ij}(\vec{k}) \left\langle V_i(\vec{k}) V_j(- \vec{k}) \right\rangle = \int \frac{\dd^3 k}{(2 \pi)^3} \mathbb{T}_{ij}(\vec{k}) (\mathbb{D}^{-1})_{ij}(\vec{k})= \int \frac{\dd^3 k}{(2 \pi)^3} \Tr(\mathbb{T} (\mathbb{D}^{-1})).
  	\end{equation}

 For the boundary part, we notice that we can rewrite it as
 \begin{equation}
   E_{Cas}^{boundary}=-\int d^4x\left(\phi K \phi\right)= -2S_{b \bar{b}}
 \end{equation}
 via partial integration and \eqref{eom1}. As such, we are brought to a quite similar expression as in \eqref{bulk1},
 	\begin{equation}
 		\frac{E_{Cas}^{boundary}}{L_1L_2} = -2 \int \frac{\dd^3 k}{(2 \pi)^3} \Tr(\mathbb{D} (\mathbb{D}^{-1})^T)
  	\end{equation}
 but given the symmetric nature of $\mathbb{D}$, see \eqref{eq:Dmatrix}, we get in dimensional regularization that
 	\begin{equation}
 		\frac{E_{Cas}^{boundary}}{L_1L_2} = -4 \int \frac{\dd^3 k}{(2 \pi)^3} =0
  	\end{equation}
 and as announced, the boundary piece of the energy-momentum tensor is not relevant for the Casimir energy\footnote{This a actually a coincidence and will not be valid in general. Including bulk interactions such as a $\lambda \phi^4$-term will already alleviate this zero contribution. }. Notice that the foregoing computation implies that the non-trivial bulk piece is actually only attributable to the $\mathbb{H}(\vec{k})$ matrix since the term in $\mathbb{D}(\vec{k})$ does lead to a contribution proportional to $\int d^3k=0$, as we have just shown.

 At last, again introducing spherical coordinates and performing the angular integration, we end up with
  	  	\begin{eqnarray}
  		\frac{E_{Cas}(q,\bar q)}{L_1L_2} &=& - \frac{1}{6 \pi^2} \int_0^{\infty} \dd r \, r^3 \frac{g \bar{g} (g - \bar{g} + g \bar{g} L) - r^2 (-g + \bar{g} + (g^2 + \bar{g}^2)L) + r^4 L}{(g - r) [e^{2Lr} (g -r )(\bar{g} + r)^2 + (g + r)(r^2 - \bar{g}^2)]}\nonumber\\
  &=& \frac{1}{12 \pi^2 L^3} \int_0^{\infty} \dd x \, x^3 e^{-x} \frac{-q \bar{q}(q - \bar{q} + q \bar{q}) + (\bar{q} - q + q^2 + \bar{q}^2) x^2 - x^4}{(q - x)(\bar{q} + x) [(q - \bar{q}) x \cosh(x) + (q \bar{q} - x^2) \sinh(x)]}.\label{res2}
  	\end{eqnarray}
 Returning to our earlier result \eqref{res1}, one can check that a partial integration in \eqref{res1}, written as $\int \dd x (f \dd g)$ and identifying $f=x^2$ and $\dd g=-2x + \ln(e^{2x} - \frac{(q + x)(x - \bar{q})}{(x - q)(\bar{q} + x)})$, nicely reduces the integral\footnote{The boundary contributions at $x\to0$ as well as at $x\to\infty$ go to zero.} to that one of \eqref{res2}, thereby showing full equivalence between the two different computation methods, and, as discussed before, also with \cite{deAlbuquerque:2003qbk}.	


\section{Discussion}
Given that we computed the Casimir energy directly from the path integral in Sec. \ref{sec:functional-integral}, one might wonder why it is useful to consider yet another computational scheme in Sec. \ref{sec:T00}, this time based on the energy-momentum tensor. There are nonetheless a few reasons to do so: (i) the energy-momentum tensor incorporates the formal local energy density (Hamiltonian) operator, which after all is the most relevant quantity for computing energy-related quantities; (ii) in case of non-planar boundaries, renormalization\footnote{It deserves mentioning that renormalization in general becomes more subtle when boundaries are present, see \cite{Borji:2022kuw,Borji:2023eyp} for a formal study for $\lambda\phi^4$ theory on a half-space.} can become more cumbersome and it is not a priori clear whether or not all computational schemes will give equal results, and if not, which one would be the physically correct one \cite{Plunien:1986ca}; (iii) for the dynamical Casimir effect, that is with time-dependent boundary conditions, a computation like in Section~II of this paper looks infeasible\footnote{It would be similar to computing the path integral in closed form when the action is perturbed with a time dependent source.}, unless for the case of small deviations w.r.t.~the static setup, when a perturbative approach in the deformation becomes applicable \cite{Golestanian:1998bx}.  But for example the case of uniformly moving plates considered in \cite{Bordag:1985rb} already falls beyond such approach.  In general, the most suitable method is then to integrate the (local, and perhaps explicitly coordinate dependent) energy density between the plates' locations\footnote{An alternative for some cases could also be to work in a more canonical (oscillator) approach and Bogoliubov transformations, see e.g.~\cite{Mintz:2006yz,Rego:2013efa}.}. We discussed the extra boundary pieces in the energy-momentum tensor due to the auxiliary fields that implement the boundary conditions. Accidentally, these boundary terms have a vanishing vacuum expectation value for the here considered case. However, this is \emph{not} generally true. In \cite{Dudal:2024xdu}, the energy-momentum tensor boundary terms were crucial in the QED case to yield the correct (electric-magnetic duality invariant) results when the most general combination of perfect electric-magnetic boundary conditions is used on the plates, see also \cite{Rode:2017yqy,Canfora:2022xcx}.

\section*{Acknowledgments}
The authors thank Fabricio Barone for pointing out relevant references. The work of D.D.~and T.O.~was supported by KU Leuven IF project C14/21/087. The work of S.S.~was funded by FWO PhD-fellowship fundamental research (file number: 1132823N). The work of B.W.M. is supported by the Brazilian agencies CNPq, CAPES, and FAPERJ. This study was financed in part by the Coordenação de Aperfeiçoamento de Pessoal de Nível Superior – Brasil (CAPES) – Finance Code 001.

\appendix
\section{Some useful (regularized) integrals}
\begin{eqnarray}
 	  	 		 \int \frac{\dd k_z}{2 \pi} \frac{1}{k^2} &=& \frac{1}{2 |\vec{k}|}\,,\qquad
 	 		 \int \frac{\dd k_z}{2 \pi} \frac{k_z^2}{k^2} = - \frac{1}{2} |\vec{k}|\,,\qquad
 	 		 \int \frac{\dd k_z}{2 \pi} \frac{1}{k^2} e^{\pm i k_z L} = \frac{1}{2 |\vec{k}|} e^{- |\vec{k}| L}\,,\qquad
 	 		 \int \frac{\dd k_z}{2 \pi} \frac{k_z}{k^2} e^{\pm i k_z L} = \pm \frac{i}{2} e^{- |\vec{k}| L}\,,\nonumber\\
 	 		 \int \frac{\dd k_z}{2 \pi} \frac{k_z^2}{k^2} e^{\pm i k_z L} &=& - \frac{1}{2} |\vec{k}| e^{- |\vec{k}| L}\,,
 \qquad	 			\int \frac{\dd k_z}{2 \pi} \frac{1}{k^4} = \frac{1}{4 |\vec{k}|^3}\,,\qquad
		\int \frac{\dd k_z}{2 \pi} \frac{k_z^2}{k^4} = \frac{1}{4 |\vec{k}|}.
 	 \end{eqnarray}

\bibliographystyle{apsrev4-2}
\bibliography{bibfile}

\begin{thebibliography}{34}%
\makeatletter
\providecommand \@ifxundefined [1]{%
 \@ifx{#1\undefined}
}%
\providecommand \@ifnum [1]{%
 \ifnum #1\expandafter \@firstoftwo
 \else \expandafter \@secondoftwo
 \fi
}%
\providecommand \@ifx [1]{%
 \ifx #1\expandafter \@firstoftwo
 \else \expandafter \@secondoftwo
 \fi
}%
\providecommand \natexlab [1]{#1}%
\providecommand \enquote  [1]{``#1''}%
\providecommand \bibnamefont  [1]{#1}%
\providecommand \bibfnamefont [1]{#1}%
\providecommand \citenamefont [1]{#1}%
\providecommand \href@noop [0]{\@secondoftwo}%
\providecommand \href [0]{\begingroup \@sanitize@url \@href}%
\providecommand \@href[1]{\@@startlink{#1}\@@href}%
\providecommand \@@href[1]{\endgroup#1\@@endlink}%
\providecommand \@sanitize@url [0]{\catcode `\\12\catcode `\$12\catcode
  `\&12\catcode `\#12\catcode `\^12\catcode `\_12\catcode `\%12\relax}%
\providecommand \@@startlink[1]{}%
\providecommand \@@endlink[0]{}%
\providecommand \url  [0]{\begingroup\@sanitize@url \@url }%
\providecommand \@url [1]{\endgroup\@href {#1}{\urlprefix }}%
\providecommand \urlprefix  [0]{URL }%
\providecommand \Eprint [0]{\href }%
\providecommand \doibase [0]{https://doi.org/}%
\providecommand \selectlanguage [0]{\@gobble}%
\providecommand \bibinfo  [0]{\@secondoftwo}%
\providecommand \bibfield  [0]{\@secondoftwo}%
\providecommand \translation [1]{[#1]}%
\providecommand \BibitemOpen [0]{}%
\providecommand \bibitemStop [0]{}%
\providecommand \bibitemNoStop [0]{.\EOS\space}%
\providecommand \EOS [0]{\spacefactor3000\relax}%
\providecommand \BibitemShut  [1]{\csname bibitem#1\endcsname}%
\let\auto@bib@innerbib\@empty
\bibitem [{\citenamefont {Casimir}\ and\ \citenamefont
  {Polder}(1948)}]{Casimir1948influence}%
  \BibitemOpen
  \bibfield  {author} {\bibinfo {author} {\bibfnamefont {H.~B.}\ \bibnamefont
  {Casimir}}\ and\ \bibinfo {author} {\bibfnamefont {D.}~\bibnamefont
  {Polder}},\ }\href@noop {} {\bibfield  {journal} {\bibinfo  {journal}
  {Physical Review}\ }\textbf {\bibinfo {volume} {73}},\ \bibinfo {pages} {360}
  (\bibinfo {year} {1948})}\BibitemShut {NoStop}%
\bibitem [{\citenamefont {Casimir}(1948)}]{Casimir1948TwoPlates}%
  \BibitemOpen
  \bibfield  {author} {\bibinfo {author} {\bibfnamefont {H.~B.~G.}\
  \bibnamefont {Casimir}},\ }\href@noop {} {\bibfield  {journal} {\bibinfo
  {journal} {Indag. Math.}\ }\textbf {\bibinfo {volume} {10}},\ \bibinfo
  {pages} {261} (\bibinfo {year} {1948})}\BibitemShut {NoStop}%
\bibitem [{\citenamefont {Bimonte}\ \emph {et~al.}(2021)\citenamefont
  {Bimonte}, \citenamefont {Spreng}, \citenamefont {Maia~Neto}, \citenamefont
  {Ingold}, \citenamefont {Klimchitskaya}, \citenamefont {Mostepanenko},\ and\
  \citenamefont {Decca}}]{Bimonte2021measurement}%
  \BibitemOpen
  \bibfield  {author} {\bibinfo {author} {\bibfnamefont {G.}~\bibnamefont
  {Bimonte}}, \bibinfo {author} {\bibfnamefont {B.}~\bibnamefont {Spreng}},
  \bibinfo {author} {\bibfnamefont {P.~A.}\ \bibnamefont {Maia~Neto}}, \bibinfo
  {author} {\bibfnamefont {G.-L.}\ \bibnamefont {Ingold}}, \bibinfo {author}
  {\bibfnamefont {G.~L.}\ \bibnamefont {Klimchitskaya}}, \bibinfo {author}
  {\bibfnamefont {V.~M.}\ \bibnamefont {Mostepanenko}},\ and\ \bibinfo {author}
  {\bibfnamefont {R.~S.}\ \bibnamefont {Decca}},\ }\href@noop {} {\bibfield
  {journal} {\bibinfo  {journal} {Universe}\ }\textbf {\bibinfo {volume} {7}},\
  \bibinfo {pages} {93} (\bibinfo {year} {2021})}\BibitemShut {NoStop}%
\bibitem [{\citenamefont {Plunien}\ \emph {et~al.}(1986)\citenamefont
  {Plunien}, \citenamefont {Muller},\ and\ \citenamefont
  {Greiner}}]{Plunien:1986ca}%
  \BibitemOpen
  \bibfield  {author} {\bibinfo {author} {\bibfnamefont {G.}~\bibnamefont
  {Plunien}}, \bibinfo {author} {\bibfnamefont {B.}~\bibnamefont {Muller}},\
  and\ \bibinfo {author} {\bibfnamefont {W.}~\bibnamefont {Greiner}},\ }\href
  {https://doi.org/10.1016/0370-1573(86)90020-7} {\bibfield  {journal}
  {\bibinfo  {journal} {Phys. Rept.}\ }\textbf {\bibinfo {volume} {134}},\
  \bibinfo {pages} {87} (\bibinfo {year} {1986})}\BibitemShut {NoStop}%
\bibitem [{\citenamefont {Bordag}\ \emph {et~al.}(2001)\citenamefont {Bordag},
  \citenamefont {Mohideen},\ and\ \citenamefont
  {Mostepanenko}}]{Bordag:2001qi}%
  \BibitemOpen
  \bibfield  {author} {\bibinfo {author} {\bibfnamefont {M.}~\bibnamefont
  {Bordag}}, \bibinfo {author} {\bibfnamefont {U.}~\bibnamefont {Mohideen}},\
  and\ \bibinfo {author} {\bibfnamefont {V.~M.}\ \bibnamefont {Mostepanenko}},\
  }\href {https://doi.org/10.1016/S0370-1573(01)00015-1} {\bibfield  {journal}
  {\bibinfo  {journal} {Phys. Rept.}\ }\textbf {\bibinfo {volume} {353}},\
  \bibinfo {pages} {1} (\bibinfo {year} {2001})},\ \Eprint
  {https://arxiv.org/abs/quant-ph/0106045} {arXiv:quant-ph/0106045}
  \BibitemShut {NoStop}%
\bibitem [{\citenamefont {Bordag}\ \emph {et~al.}(2009)\citenamefont {Bordag},
  \citenamefont {Klimchitskaya}, \citenamefont {Mohideen},\ and\ \citenamefont
  {Mostepanenko}}]{Bordag:2009zz}%
  \BibitemOpen
  \bibfield  {author} {\bibinfo {author} {\bibfnamefont {M.}~\bibnamefont
  {Bordag}}, \bibinfo {author} {\bibfnamefont {G.~L.}\ \bibnamefont
  {Klimchitskaya}}, \bibinfo {author} {\bibfnamefont {U.}~\bibnamefont
  {Mohideen}},\ and\ \bibinfo {author} {\bibfnamefont {V.~M.}\ \bibnamefont
  {Mostepanenko}},\ }\href@noop {} {\emph {\bibinfo {title} {{Advances in the
  Casimir effect}}}},\ Vol.\ \bibinfo {volume} {145}\ (\bibinfo  {publisher}
  {Oxford University Press},\ \bibinfo {year} {2009})\BibitemShut {NoStop}%
\bibitem [{\citenamefont {Milton}(2004)}]{Milton:2004ya}%
  \BibitemOpen
  \bibfield  {author} {\bibinfo {author} {\bibfnamefont {K.~A.}\ \bibnamefont
  {Milton}},\ }\href {https://doi.org/10.1088/0305-4470/37/38/R01} {\bibfield
  {journal} {\bibinfo  {journal} {J. Phys. A}\ }\textbf {\bibinfo {volume}
  {37}},\ \bibinfo {pages} {R209} (\bibinfo {year} {2004})},\ \Eprint
  {https://arxiv.org/abs/hep-th/0406024} {arXiv:hep-th/0406024} \BibitemShut
  {NoStop}%
\bibitem [{\citenamefont {Zou}\ \emph {et~al.}(2013)\citenamefont {Zou},
  \citenamefont {Marcet}, \citenamefont {Rodriguez}, \citenamefont {Reid},
  \citenamefont {McCauley}, \citenamefont {Kravchenko}, \citenamefont {Lu},
  \citenamefont {Bao}, \citenamefont {Johnson},\ and\ \citenamefont
  {Chan}}]{Zou2013casimir}%
  \BibitemOpen
  \bibfield  {author} {\bibinfo {author} {\bibfnamefont {J.}~\bibnamefont
  {Zou}}, \bibinfo {author} {\bibfnamefont {Z.}~\bibnamefont {Marcet}},
  \bibinfo {author} {\bibfnamefont {A.~W.}\ \bibnamefont {Rodriguez}}, \bibinfo
  {author} {\bibfnamefont {M.~H.}\ \bibnamefont {Reid}}, \bibinfo {author}
  {\bibfnamefont {A.~P.}\ \bibnamefont {McCauley}}, \bibinfo {author}
  {\bibfnamefont {I.~I.}\ \bibnamefont {Kravchenko}}, \bibinfo {author}
  {\bibfnamefont {T.}~\bibnamefont {Lu}}, \bibinfo {author} {\bibfnamefont
  {Y.}~\bibnamefont {Bao}}, \bibinfo {author} {\bibfnamefont {S.~G.}\
  \bibnamefont {Johnson}},\ and\ \bibinfo {author} {\bibfnamefont {H.~B.}\
  \bibnamefont {Chan}},\ }\href@noop {} {\bibfield  {journal} {\bibinfo
  {journal} {Nature communications}\ }\textbf {\bibinfo {volume} {4}},\
  \bibinfo {pages} {1845} (\bibinfo {year} {2013})}\BibitemShut {NoStop}%
\bibitem [{\citenamefont {Chernodub}\ \emph {et~al.}(2018)\citenamefont
  {Chernodub}, \citenamefont {Goy}, \citenamefont {Molochkov},\ and\
  \citenamefont {Nguyen}}]{Chernodub2018casimir}%
  \BibitemOpen
  \bibfield  {author} {\bibinfo {author} {\bibfnamefont {M.}~\bibnamefont
  {Chernodub}}, \bibinfo {author} {\bibfnamefont {V.}~\bibnamefont {Goy}},
  \bibinfo {author} {\bibfnamefont {A.}~\bibnamefont {Molochkov}},\ and\
  \bibinfo {author} {\bibfnamefont {H.~H.}\ \bibnamefont {Nguyen}},\
  }\href@noop {} {\bibfield  {journal} {\bibinfo  {journal} {Physical Review
  Letters}\ }\textbf {\bibinfo {volume} {121}},\ \bibinfo {pages} {191601}
  (\bibinfo {year} {2018})}\BibitemShut {NoStop}%
\bibitem [{\citenamefont {Chernodub}\ \emph {et~al.}(2023)\citenamefont
  {Chernodub}, \citenamefont {Goy}, \citenamefont {Molochkov},\ and\
  \citenamefont {Tanashkin}}]{Chernodub:2023dok}%
  \BibitemOpen
  \bibfield  {author} {\bibinfo {author} {\bibfnamefont {M.~N.}\ \bibnamefont
  {Chernodub}}, \bibinfo {author} {\bibfnamefont {V.~A.}\ \bibnamefont {Goy}},
  \bibinfo {author} {\bibfnamefont {A.~V.}\ \bibnamefont {Molochkov}},\ and\
  \bibinfo {author} {\bibfnamefont {A.~S.}\ \bibnamefont {Tanashkin}},\ }\href
  {https://doi.org/10.1103/PhysRevD.108.014515} {\bibfield  {journal} {\bibinfo
   {journal} {Phys. Rev. D}\ }\textbf {\bibinfo {volume} {108}},\ \bibinfo
  {pages} {014515} (\bibinfo {year} {2023})},\ \Eprint
  {https://arxiv.org/abs/2302.00376} {arXiv:2302.00376 [hep-lat]} \BibitemShut
  {NoStop}%
\bibitem [{\citenamefont {Asorey}\ \emph {et~al.}(2024)\citenamefont {Asorey},
  \citenamefont {Iuliano},\ and\ \citenamefont {Ezquerro}}]{Asorey:2024vaw}%
  \BibitemOpen
  \bibfield  {author} {\bibinfo {author} {\bibfnamefont {M.}~\bibnamefont
  {Asorey}}, \bibinfo {author} {\bibfnamefont {C.}~\bibnamefont {Iuliano}},\
  and\ \bibinfo {author} {\bibfnamefont {F.}~\bibnamefont {Ezquerro}},\ }\href
  {https://doi.org/10.3390/physics6020040} {\bibfield  {journal} {\bibinfo
  {journal} {Physics}\ }\textbf {\bibinfo {volume} {6}},\ \bibinfo {pages}
  {613} (\bibinfo {year} {2024})},\ \Eprint {https://arxiv.org/abs/2405.03768}
  {arXiv:2405.03768 [hep-th]} \BibitemShut {NoStop}%
\bibitem [{\citenamefont {Karabali}\ and\ \citenamefont
  {Nair}(2018)}]{Karabali:2018ael}%
  \BibitemOpen
  \bibfield  {author} {\bibinfo {author} {\bibfnamefont {D.}~\bibnamefont
  {Karabali}}\ and\ \bibinfo {author} {\bibfnamefont {V.~P.}\ \bibnamefont
  {Nair}},\ }\href {https://doi.org/10.1103/PhysRevD.98.105009} {\bibfield
  {journal} {\bibinfo  {journal} {Phys. Rev. D}\ }\textbf {\bibinfo {volume}
  {98}},\ \bibinfo {pages} {105009} (\bibinfo {year} {2018})},\ \Eprint
  {https://arxiv.org/abs/1808.07979} {arXiv:1808.07979 [hep-th]} \BibitemShut
  {NoStop}%
\bibitem [{\citenamefont {Bordag}\ \emph {et~al.}(1985)\citenamefont {Bordag},
  \citenamefont {Robaschik},\ and\ \citenamefont {Wieczorek}}]{Bordag:1983zk}%
  \BibitemOpen
  \bibfield  {author} {\bibinfo {author} {\bibfnamefont {M.}~\bibnamefont
  {Bordag}}, \bibinfo {author} {\bibfnamefont {D.}~\bibnamefont {Robaschik}},\
  and\ \bibinfo {author} {\bibfnamefont {E.}~\bibnamefont {Wieczorek}},\ }\href
  {https://doi.org/10.1016/S0003-4916(85)80009-9} {\bibfield  {journal}
  {\bibinfo  {journal} {Annals Phys.}\ }\textbf {\bibinfo {volume} {165}},\
  \bibinfo {pages} {192} (\bibinfo {year} {1985})}\BibitemShut {NoStop}%
\bibitem [{\citenamefont {Dudal}\ \emph {et~al.}(2020)\citenamefont {Dudal},
  \citenamefont {Pais},\ and\ \citenamefont {Rosa}}]{Dudal:2020yah}%
  \BibitemOpen
  \bibfield  {author} {\bibinfo {author} {\bibfnamefont {D.}~\bibnamefont
  {Dudal}}, \bibinfo {author} {\bibfnamefont {P.}~\bibnamefont {Pais}},\ and\
  \bibinfo {author} {\bibfnamefont {L.}~\bibnamefont {Rosa}},\ }\href
  {https://doi.org/10.1103/PhysRevD.102.016026} {\bibfield  {journal} {\bibinfo
   {journal} {Phys. Rev. D}\ }\textbf {\bibinfo {volume} {102}},\ \bibinfo
  {pages} {016026} (\bibinfo {year} {2020})},\ \Eprint
  {https://arxiv.org/abs/2005.12693} {arXiv:2005.12693 [hep-th]} \BibitemShut
  {NoStop}%
\bibitem [{\citenamefont {Canfora}\ \emph {et~al.}(2022)\citenamefont
  {Canfora}, \citenamefont {Dudal}, \citenamefont {Oosthuyse}, \citenamefont
  {Pais},\ and\ \citenamefont {Rosa}}]{Canfora:2022xcx}%
  \BibitemOpen
  \bibfield  {author} {\bibinfo {author} {\bibfnamefont {F.}~\bibnamefont
  {Canfora}}, \bibinfo {author} {\bibfnamefont {D.}~\bibnamefont {Dudal}},
  \bibinfo {author} {\bibfnamefont {T.}~\bibnamefont {Oosthuyse}}, \bibinfo
  {author} {\bibfnamefont {P.}~\bibnamefont {Pais}},\ and\ \bibinfo {author}
  {\bibfnamefont {L.}~\bibnamefont {Rosa}},\ }\href
  {https://doi.org/10.1007/JHEP09(2022)095} {\bibfield  {journal} {\bibinfo
  {journal} {JHEP}\ }\textbf {\bibinfo {volume} {09}},\ \bibinfo {pages}
  {095}},\ \Eprint {https://arxiv.org/abs/2207.09175} {arXiv:2207.09175
  [hep-th]} \BibitemShut {NoStop}%
\bibitem [{\citenamefont {Oosthuyse}\ and\ \citenamefont
  {Dudal}(2023)}]{Oosthuyse:2023mbs}%
  \BibitemOpen
  \bibfield  {author} {\bibinfo {author} {\bibfnamefont {T.}~\bibnamefont
  {Oosthuyse}}\ and\ \bibinfo {author} {\bibfnamefont {D.}~\bibnamefont
  {Dudal}},\ }\href {https://doi.org/10.21468/SciPostPhys.15.5.213} {\bibfield
  {journal} {\bibinfo  {journal} {SciPost Phys.}\ }\textbf {\bibinfo {volume}
  {15}},\ \bibinfo {pages} {213} (\bibinfo {year} {2023})},\ \Eprint
  {https://arxiv.org/abs/2301.12870} {arXiv:2301.12870 [hep-th]} \BibitemShut
  {NoStop}%
\bibitem [{\citenamefont {Golestanian}\ and\ \citenamefont
  {Kardar}(1998)}]{Golestanian:1998bx}%
  \BibitemOpen
  \bibfield  {author} {\bibinfo {author} {\bibfnamefont {R.}~\bibnamefont
  {Golestanian}}\ and\ \bibinfo {author} {\bibfnamefont {M.}~\bibnamefont
  {Kardar}},\ }\href {https://doi.org/10.1103/PhysRevA.58.1713} {\bibfield
  {journal} {\bibinfo  {journal} {Phys. Rev. A}\ }\textbf {\bibinfo {volume}
  {58}},\ \bibinfo {pages} {1713} (\bibinfo {year} {1998})},\ \Eprint
  {https://arxiv.org/abs/quant-ph/9802017} {arXiv:quant-ph/9802017}
  \BibitemShut {NoStop}%
\bibitem [{\citenamefont {de~Albuquerque}\ and\ \citenamefont
  {Cavalcanti}(2004)}]{deAlbuquerque:2003qbk}%
  \BibitemOpen
  \bibfield  {author} {\bibinfo {author} {\bibfnamefont {L.~C.}\ \bibnamefont
  {de~Albuquerque}}\ and\ \bibinfo {author} {\bibfnamefont {R.~M.}\
  \bibnamefont {Cavalcanti}},\ }\href
  {https://doi.org/10.1088/0305-4470/37/27/012} {\bibfield  {journal} {\bibinfo
   {journal} {J. Phys. A}\ }\textbf {\bibinfo {volume} {37}},\ \bibinfo {pages}
  {7039} (\bibinfo {year} {2004})},\ \Eprint
  {https://arxiv.org/abs/hep-th/0311052} {arXiv:hep-th/0311052} \BibitemShut
  {NoStop}%
\bibitem [{\citenamefont {Ambjorn}\ and\ \citenamefont
  {Wolfram}(1983)}]{Ambjorn:1981xw}%
  \BibitemOpen
  \bibfield  {author} {\bibinfo {author} {\bibfnamefont {J.}~\bibnamefont
  {Ambjorn}}\ and\ \bibinfo {author} {\bibfnamefont {S.}~\bibnamefont
  {Wolfram}},\ }\href {https://doi.org/10.1016/0003-4916(83)90065-9} {\bibfield
   {journal} {\bibinfo  {journal} {Annals Phys.}\ }\textbf {\bibinfo {volume}
  {147}},\ \bibinfo {pages} {1} (\bibinfo {year} {1983})}\BibitemShut {NoStop}%
\bibitem [{\citenamefont {Brown}\ and\ \citenamefont
  {Maclay}(1969)}]{BrownMaclay1969}%
  \BibitemOpen
  \bibfield  {author} {\bibinfo {author} {\bibfnamefont {L.~S.}\ \bibnamefont
  {Brown}}\ and\ \bibinfo {author} {\bibfnamefont {G.~J.}\ \bibnamefont
  {Maclay}},\ }\href {https://doi.org/10.1103/PhysRev.184.1272} {\bibfield
  {journal} {\bibinfo  {journal} {Phys. Rev.}\ }\textbf {\bibinfo {volume}
  {184}},\ \bibinfo {pages} {1272} (\bibinfo {year} {1969})}\BibitemShut
  {NoStop}%
\bibitem [{\citenamefont {Barone}(2003)}]{Barone:2003crf}%
  \BibitemOpen
  \bibfield  {author} {\bibinfo {author} {\bibfnamefont {F.~A.}\ \bibnamefont
  {Barone}},\ }\emph {\bibinfo {title} {{Corre\c{c}\~oes radiativas em teoria
  qu\^antica de campos sob condi\c{c}\~oes de contorno}}},\ \href@noop {}
  {Ph.D. thesis},\ \bibinfo  {school} {Rio de Janeiro Federal U.} (\bibinfo
  {year} {2003})\BibitemShut {NoStop}%
\bibitem [{\citenamefont {{Albuquerque}}\ \emph {et~al.}(1997)\citenamefont
  {{Albuquerque}}, \citenamefont {{Farina}},\ and\ \citenamefont
  {{Theodoro}}}]{Albuquerque:1997BJP}%
  \BibitemOpen
  \bibfield  {author} {\bibinfo {author} {\bibfnamefont {L.~C.}\ \bibnamefont
  {{Albuquerque}}}, \bibinfo {author} {\bibfnamefont {C.}~\bibnamefont
  {{Farina}}},\ and\ \bibinfo {author} {\bibfnamefont {L.~G.~A.}\ \bibnamefont
  {{Theodoro}}},\ }\href {https://doi.org/10.1590/S0103-97331997000400010}
  {\bibfield  {journal} {\bibinfo  {journal} {Brazilian Journal of Physics}\
  }\textbf {\bibinfo {volume} {27}},\ \bibinfo {pages} {488} (\bibinfo {year}
  {1997})}\BibitemShut {NoStop}%
\bibitem [{\citenamefont {Barone}\ \emph {et~al.}(2016)\citenamefont {Barone},
  \citenamefont {Nogueira},\ and\ \citenamefont {Pimentel}}]{Barone:2016rnf}%
  \BibitemOpen
  \bibfield  {author} {\bibinfo {author} {\bibfnamefont {F.~A.}\ \bibnamefont
  {Barone}}, \bibinfo {author} {\bibfnamefont {A.~A.}\ \bibnamefont
  {Nogueira}},\ and\ \bibinfo {author} {\bibfnamefont {B.~M.}\ \bibnamefont
  {Pimentel}},\ }\href {https://doi.org/10.1590/1806-9126-RBEF-2016-0033}
  {\bibfield  {journal} {\bibinfo  {journal} {Rev. Bras. Ens. Fis.}\ }\textbf
  {\bibinfo {volume} {38}},\ \bibinfo {pages} {e3317} (\bibinfo {year}
  {2016})}\BibitemShut {NoStop}%
\bibitem [{\citenamefont {Belinfante}(1940)}]{Belinfante1940OnTC}%
  \BibitemOpen
  \bibfield  {author} {\bibinfo {author} {\bibfnamefont {F.~J.}\ \bibnamefont
  {Belinfante}},\ }\href {https://doi.org/10.1016/S0031-8914(40)90091-X}
  {\bibfield  {journal} {\bibinfo  {journal} {Physica D: Nonlinear Phenomena}\
  }\textbf {\bibinfo {volume} {7}},\ \bibinfo {pages} {449} (\bibinfo {year}
  {1940})}\BibitemShut {NoStop}%
\bibitem [{\citenamefont {Baker}\ \emph {et~al.}(2021)\citenamefont {Baker},
  \citenamefont {Linnemann},\ and\ \citenamefont {Smeenk}}]{Baker:2021hly}%
  \BibitemOpen
  \bibfield  {author} {\bibinfo {author} {\bibfnamefont {M.~R.}\ \bibnamefont
  {Baker}}, \bibinfo {author} {\bibfnamefont {N.}~\bibnamefont {Linnemann}},\
  and\ \bibinfo {author} {\bibfnamefont {C.}~\bibnamefont {Smeenk}}\ }\href
  {https://doi.org/10.48550/arXiv.2107.10329} {10.48550/arXiv.2107.10329}
  (\bibinfo {year} {2021}),\ \Eprint {https://arxiv.org/abs/2107.10329}
  {arXiv:2107.10329 [physics.hist-ph]} \BibitemShut {NoStop}%
\bibitem [{\citenamefont {Freese}(2022)}]{Freese_2022}%
  \BibitemOpen
  \bibfield  {author} {\bibinfo {author} {\bibfnamefont {A.}~\bibnamefont
  {Freese}},\ }\href {https://doi.org/10.1103/PhysRevD.106.125012} {\bibfield
  {journal} {\bibinfo  {journal} {Phys. Rev. D}\ }\textbf {\bibinfo {volume}
  {106}},\ \bibinfo {pages} {125012} (\bibinfo {year} {2022})}\BibitemShut
  {NoStop}%
\bibitem [{\citenamefont {Romeo}\ and\ \citenamefont
  {Saharian}(2002)}]{Romeo:2000wt}%
  \BibitemOpen
  \bibfield  {author} {\bibinfo {author} {\bibfnamefont {A.}~\bibnamefont
  {Romeo}}\ and\ \bibinfo {author} {\bibfnamefont {A.~A.}\ \bibnamefont
  {Saharian}},\ }\href {https://doi.org/10.1088/0305-4470/35/5/312} {\bibfield
  {journal} {\bibinfo  {journal} {J. Phys. A}\ }\textbf {\bibinfo {volume}
  {35}},\ \bibinfo {pages} {1297} (\bibinfo {year} {2002})},\ \Eprint
  {https://arxiv.org/abs/hep-th/0007242} {arXiv:hep-th/0007242} \BibitemShut
  {NoStop}%
\bibitem [{\citenamefont {Borji}\ and\ \citenamefont
  {Kopper}(2022)}]{Borji:2022kuw}%
  \BibitemOpen
  \bibfield  {author} {\bibinfo {author} {\bibfnamefont {M.}~\bibnamefont
  {Borji}}\ and\ \bibinfo {author} {\bibfnamefont {C.}~\bibnamefont {Kopper}},\
  }\href {https://doi.org/10.1063/5.0097164} {\bibfield  {journal} {\bibinfo
  {journal} {J. Math. Phys.}\ }\textbf {\bibinfo {volume} {63}},\ \bibinfo
  {pages} {092304} (\bibinfo {year} {2022})},\ \Eprint
  {https://arxiv.org/abs/2204.04326} {arXiv:2204.04326 [math-ph]} \BibitemShut
  {NoStop}%
\bibitem [{\citenamefont {Borji}\ and\ \citenamefont
  {Kopper}(2024)}]{Borji:2023eyp}%
  \BibitemOpen
  \bibfield  {author} {\bibinfo {author} {\bibfnamefont {M.}~\bibnamefont
  {Borji}}\ and\ \bibinfo {author} {\bibfnamefont {C.}~\bibnamefont {Kopper}},\
  }\href {https://doi.org/10.1063/5.0164178} {\bibfield  {journal} {\bibinfo
  {journal} {J. Math. Phys.}\ }\textbf {\bibinfo {volume} {65}},\ \bibinfo
  {pages} {022301} (\bibinfo {year} {2024})},\ \Eprint
  {https://arxiv.org/abs/2305.18862} {arXiv:2305.18862 [math-ph]} \BibitemShut
  {NoStop}%
\bibitem [{\citenamefont {Bordag}\ \emph {et~al.}(1986)\citenamefont {Bordag},
  \citenamefont {Dittes},\ and\ \citenamefont {Robaschik}}]{Bordag:1985rb}%
  \BibitemOpen
  \bibfield  {author} {\bibinfo {author} {\bibfnamefont {M.}~\bibnamefont
  {Bordag}}, \bibinfo {author} {\bibfnamefont {F.~M.}\ \bibnamefont {Dittes}},\
  and\ \bibinfo {author} {\bibfnamefont {D.}~\bibnamefont {Robaschik}},\
  }\href@noop {} {\bibfield  {journal} {\bibinfo  {journal} {Sov. J. Nucl.
  Phys.}\ }\textbf {\bibinfo {volume} {43}},\ \bibinfo {pages} {1034} (\bibinfo
  {year} {1986})}\BibitemShut {NoStop}%
\bibitem [{\citenamefont {Mintz}\ \emph {et~al.}(2006)\citenamefont {Mintz},
  \citenamefont {Farina}, \citenamefont {Maia~Neto},\ and\ \citenamefont
  {Rodrigues}}]{Mintz:2006yz}%
  \BibitemOpen
  \bibfield  {author} {\bibinfo {author} {\bibfnamefont {B.}~\bibnamefont
  {Mintz}}, \bibinfo {author} {\bibfnamefont {C.}~\bibnamefont {Farina}},
  \bibinfo {author} {\bibfnamefont {P.~A.}\ \bibnamefont {Maia~Neto}},\ and\
  \bibinfo {author} {\bibfnamefont {R.~B.}\ \bibnamefont {Rodrigues}},\ }\href
  {https://doi.org/10.1088/0305-4470/39/36/013} {\bibfield  {journal} {\bibinfo
   {journal} {J. Phys. A}\ }\textbf {\bibinfo {volume} {39}},\ \bibinfo {pages}
  {11325} (\bibinfo {year} {2006})},\ \Eprint
  {https://arxiv.org/abs/hep-th/0605221} {arXiv:hep-th/0605221} \BibitemShut
  {NoStop}%
\bibitem [{\citenamefont {Rego}\ \emph {et~al.}(2013)\citenamefont {Rego},
  \citenamefont {Mintz}, \citenamefont {Farina},\ and\ \citenamefont
  {Alves}}]{Rego:2013efa}%
  \BibitemOpen
  \bibfield  {author} {\bibinfo {author} {\bibfnamefont {A.~L.~C.}\
  \bibnamefont {Rego}}, \bibinfo {author} {\bibfnamefont {B.~W.}\ \bibnamefont
  {Mintz}}, \bibinfo {author} {\bibfnamefont {C.}~\bibnamefont {Farina}},\ and\
  \bibinfo {author} {\bibfnamefont {D.~T.}\ \bibnamefont {Alves}},\ }\href
  {https://doi.org/10.1103/PhysRevD.87.045024} {\bibfield  {journal} {\bibinfo
  {journal} {Phys. Rev. D}\ }\textbf {\bibinfo {volume} {87}},\ \bibinfo
  {pages} {045024} (\bibinfo {year} {2013})},\ \Eprint
  {https://arxiv.org/abs/1302.4709} {arXiv:1302.4709 [quant-ph]} \BibitemShut
  {NoStop}%
\bibitem [{\citenamefont {Dudal}\ \emph {et~al.}(2024)\citenamefont {Dudal},
  \citenamefont {Gobeyn}, \citenamefont {Oosthuyse}, \citenamefont {Stouten},\
  and\ \citenamefont {Vercauteren}}]{Dudal:2024xdu}%
  \BibitemOpen
  \bibfield  {author} {\bibinfo {author} {\bibfnamefont {D.}~\bibnamefont
  {Dudal}}, \bibinfo {author} {\bibfnamefont {A.}~\bibnamefont {Gobeyn}},
  \bibinfo {author} {\bibfnamefont {T.}~\bibnamefont {Oosthuyse}}, \bibinfo
  {author} {\bibfnamefont {S.}~\bibnamefont {Stouten}},\ and\ \bibinfo {author}
  {\bibfnamefont {D.}~\bibnamefont {Vercauteren}},\ }\href@noop {} {\
  (\bibinfo {year} {2024})},\ \Eprint {https://arxiv.org/abs/2406.19743}
  {arXiv:2406.19743 [hep-th]} \BibitemShut {NoStop}%
\bibitem [{\citenamefont {Rode}\ \emph {et~al.}(2018)\citenamefont {Rode},
  \citenamefont {Bennett},\ and\ \citenamefont {Buhmann}}]{Rode:2017yqy}%
  \BibitemOpen
  \bibfield  {author} {\bibinfo {author} {\bibfnamefont {S.}~\bibnamefont
  {Rode}}, \bibinfo {author} {\bibfnamefont {R.}~\bibnamefont {Bennett}},\ and\
  \bibinfo {author} {\bibfnamefont {S.~Y.}\ \bibnamefont {Buhmann}},\ }\href
  {https://doi.org/10.1088/1367-2630/aaaa44} {\bibfield  {journal} {\bibinfo
  {journal} {New J. Phys.}\ }\textbf {\bibinfo {volume} {20}},\ \bibinfo
  {pages} {043024} (\bibinfo {year} {2018})},\ \Eprint
  {https://arxiv.org/abs/1710.01509} {arXiv:1710.01509 [quant-ph]} \BibitemShut
  {NoStop}%
\end{thebibliography}%

\end{document}